\documentclass[
reprint,
superscriptaddress,
bibnotes,
amssymb, amsmath,
aps,
prb,
]{revtex4-2}

\usepackage{graphicx}% Include figure files
\usepackage{dcolumn}% Align table columns on decimal point
\usepackage{bm}% bold math
\usepackage[utf8]{inputenc}
\usepackage[T1]{fontenc}
\usepackage{etoolbox}
\usepackage{hyperref}
\usepackage{subfigure}
\usepackage{xcolor}
\makeatletter
\def\@email#1#2{%
 \endgroup
 \patchcmd{\titleblock@produce}
  {\frontmatter@RRAPformat} {\frontmatter@RRAPformat{\produce@RRAP{*#1\href{mailto:#2}{#2}}}\frontmatter@RRAPformat}
  {}{}
}%
\makeatother
\begin{document}

\preprint{AIP/123-QED}

\title{Control Protocol for Dynamic Synthesis of Qubit and Qudit Gates Using Photonic Pulses and Magnetic Fields
}

\author{A. F. Urquijo Rodr\'iguez}
 \altaffiliation[Correspondig author:]{afurquijor@unal.edu.co (A. F. Urquijo Rodrígez)}
 \affiliation{Grupo de Superconductividad y Nanotecnología, Departamento de Física, Universidad Nacional de Colombia, Carrera 45 No. 26-85 Bogotá, 111321, Colombia}

\author{ Edgar A. G\'omez }
 \affiliation{Grupo de Investigación en Física Teórica y Computacional, Programa de Física, Universidad del Quindío, 630004, Armenia, Colombia}

\author{H. Vinck-Posada}
\affiliation{Grupo de Superconductividad y Nanotecnología, Departamento de Física, Universidad Nacional de Colombia, Carrera 45 No. 26-85 Bogotá, 111321, Colombia}
\begin{abstract}
We propose a theoretical control protocol designed for the dynamic synthesis of single-qubit and four-level qudit quantum gates using external parameters, such as photonic Gaussian pulses and magnetic fields, in a microcavity–quantum well system. Our approach takes advantage of tunable coherent light–matter interactions that can be modulated by the magnetic field between the exciton and negative trion coupled to the lowest photonic mode. We demonstrate that it is possible to achieve precise manipulation of populations of encoded quantum states through the unitary evolution of the system. In particular, we illustrate our optimization method for generating a single-qubit gate with a mean fidelity of $99.99$ as well as the realization of an iSWAP gate in the four-level qudit case with a fidelity of $99.6$.
\end{abstract}
\maketitle
\section{\label{sec:level1}Introduction}
Exciton polaritons are quasiparticles formed from a statistical blend of cavity photons and bound electron-hole states within a mesoscopic semiconductor system. These quasiparticles exhibit intriguing properties owing to their bosonic nature~\cite{Deng2010} and the small effective masses derived from their photonic components. The most notable properties are the Bose–Einstein condensation near room temperature~\cite{Imamoglu1996,Kasprzak2006}, polariton lasing~\cite{Byrnes2014}, and vorticity in the quantum fluid regime~\cite{Kavokin2023}. Furthermore, exciton polaritons are expected to function in the quantum regime~\cite{Savasta2005}, theoretically displaying entanglement and quantum correlation signatures~\cite{Lagoudakis2008,Schwendimann2003}, which makes them promising candidates for applications in quantum computation~\cite{Ghosh2020,Anantharaman2021, Demirchyan2014,Malpuech2015} or quantum simulators. Recent studies have focused on manipulating the quantum properties of polaritonic systems by applying an external magnetic field. This manipulation influences the excitonic component of the polariton, allowing for the tuning of the exciton energy, altering the Zeeman splitting~\cite{Larinov2010,Liew2011}, and enhancing the exciton oscillator strength~\cite{Devaud2015}. The quantum characteristics of this platform can be further enhanced by considering the many-body effects induced by the Coulomb interactions within the semiconductor. This can lead to the emergence of more exotic quasiparticles, such as trions, in doped semiconductors~\cite{Devaud2009}. Trions, which are the bound states of three fermions, have also been shown to coherently interact with the light modes confined within a cavity~\cite{Rapaport2001,Rana2021}. This opens up the possibility of magnetic tuning (electrons, excitons, and trions) inside the microcavities, enabling the use of exciton–trions systems as multilevel qudits. In recent years, there has been considerable interest in generating quantum gates on semiconductor platforms and achieving quantum control using external parameters. For example, spin-photon hybrid systems, which are considered electron-spin qubits in quantum dots within optical microcavities~\cite{Wei:2014, Liu2018, Han2021}, photonic systems with more than one degree of freedom in charged quantum dots inside optical microcavities~\cite{Xia_2018}, and qubits based on conduction-band electron spins in quantum dots embedded within a microcavity~\cite{Dong2009}. Recently, researchers demonstrated the feasibility of implementing quantum gates through controlled parametric interactions between the control and target polariton modes~\cite{Kyriienko2016, Ricco2024, Savvidis2024}. However, few theoretical studies have focused on the influence of external magnetic fields and ultrashort pulse engineering on microcavity–quantum well systems and how they can be utilized to control quantum state transitions~\cite{Jimenez2017, Villas2017, Paspalakis2012}. The aim of this study is to introduce a theoretical approach for synthesizing single-qubit and four-level qubit quantum gates through the quantum evolution of a trion-exciton-polariton system utilizing external magnetic fields and a set of coherent pumping pulses as the physical control parameters. The remainder of this paper is organized as follows. In Section ~\ref{sec:teoria}, we present a theoretical model for describing a semiconductor quantum well (QW) immersed in an optical cavity within the framework of a finite system. Specifically, we examined the properties of exciton and trion states using a finite-system Hamiltonian model. In addition, we discuss the optimization algorithm for the quantum gates in detail. In Section~\ref{sec:results}, we provide a detailed analysis of the numerical results, focusing on the effects of the magnetic field on the Coulomb interaction energy, exciton and trion energies, and the synthesis of single-qubit and four-qudit gates. Finally, concluding remarks are presented in Section~\ref{sec:conclusion}. In Appendix~\ref{sec:appendix_a}, we present the mathematical details of Coulomb matrix element computation. The single-particle-state ordering used in this study is briefly discussed in Appendix~\ref{sec:ordering}.
\section{ Theoretical framework}\label{sec:teoria}
\subsection{Finite System Hamiltonian}
\begin{figure}[!ht]
        \centering
        \includegraphics[width=0.9\linewidth]{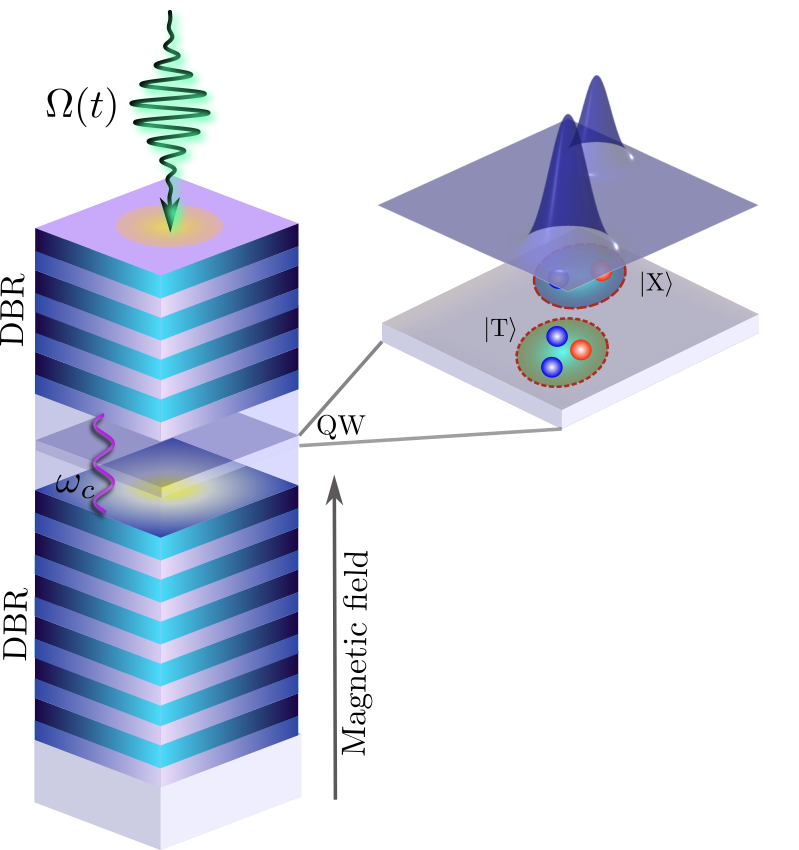}
        \caption{Illustration of the physical system: A single quantum well (QW) is immersed in a semiconductor microcavity formed by two distributed Bragg reflectors (DBR) and exposed to a perpendicular magnetic field. Within this material, trion and exciton resonances can emerge owing to the interaction between the charged carriers and the confined lowest photonic mode. For the interpretation of the colors in the figure(s), the reader is referred to the web version of this article.}
        \label{fig:figure-1}
    \end{figure}

The system considered in this study consisted of a single QW embedded in a planar semiconductor microcavity subjected to an external perpendicular magnetic field. Exciton and negative trion states can arise in semiconductor systems owing to the Coulomb correlation between the electrons and holes. Both quasiparticles can interact with the lowest photonic mode confined inside the cavity, as shown in Fig.\ref{fig:figure-1}. We limit our consideration to the tight confinement limit in quantum wells (QW) and ignore band mixing effects. Furthermore, we explicitly included the spin degree of freedom of the particle in our model to consider the optical selection rules that govern the light-matter interaction in the system. Using basis states composed of the direct product of the Landau Levels (LL) and QW lowest $z$ sub-band wave functions, we can write the system Hamiltonian in the second quantization framework as follows~\cite{Urquijo-Rodríguez_2024}:
\begin{eqnarray} \nonumber
        H &=& \sum_{i}\{ \mathcal{E}^{(e)}_{i} e^\dagger_i e_i + \mathcal{E}^{(h)}_{\overline{i}} h^\dagger_{\overline{i}} h_{\overline{i}} \} + \hbar \omega a^\dagger a \\  \nonumber
        &+& \frac{\beta}{2} \Big[ \sum_{ijkl} \langle ij|V |kl\rangle  e^\dagger_{i}e^\dagger_{j} e^\dagger_{l} e^\dagger_{k} - 2\sum_{i\overline{j}k \overline{l}} \langle i \overline{j}| V |k \overline{l} \rangle e^\dagger_{i} h^\dagger_{\overline{j}} h_{\overline{l}} e_k \Big] \\ 
        &+& \hbar g \sum_{i} \{ e^\dagger_{i\uparrow} h^{\dagger}_{\overline{i} \downarrow} a + a^\dagger h_{\overline{i}\downarrow} e_{i\uparrow} \},
        \label{eq:sys_hamiltonian}
\end{eqnarray}
where $e_j$ and $h_{\overline{j}}$ are the fermionic annihilation operators for the electrons and holes, respectively. The subscript $j$ labels the single-particle states $\{|n_j,l_j,s_j \rangle \}$, where $n_j$ is the radial quantum number, $l_j$  is the $z$-projection of the angular momentum, and $s_j$ is the $z$ component of the $j$-th spin particle. Note that the overline notation denotes the states for holes; that is, $\textstyle l_j^{(h)} = -l_j^{(e)}$ and $s_j^{(h)} = -s_j^{(e)}$. In addition, $a$ corresponds to the annihilation operator of the lowest photonic mode with left circular polarization. The Hamiltonian, described in Eq.(~\ref{eq:sys_hamiltonian}) encompasses several processes. The first part considers the single-particle energies, which incorporate the effects of the cyclotron, Zeeman, and QW sub-band energies for both electrons and holes, expressed as follows:
\begin{eqnarray} \nonumber
        \mathcal{E}^{e(h)}_{i} &=& \frac{\hbar \omega^{e(h)}_c}{2}( 2 n_i + |l_i| \pm l_i + 1) \pm g_{e(h)}\mu_{B} B s_i \\
        &+& \frac{\hbar^2 \pi^2}{2 m_{e(h)} L} + E_{g}.
\end{eqnarray}
In this context, $E_g$ represents the energy gap, whereas $\omega^{e(h)}_c = eB/m_{e(h)}$ represents the cyclotron frequency of electrons (holes). The effective gyromagnetic factor for electrons (holes) is denoted by $g_{e(h)}$ and $\mu_B$ denotes the Bohr magneton. The width of the quantum well is denoted by $L$. The masses of the electrons and holes are represented by $m_e$ and $m_h$, respectively. The second part of the Hamiltonian accounts for the Coulomb interaction between electrons and holes. Importantly, this interaction term is influenced by the magnetic field through the Coulomb amplitude strength $\beta$ and the matrix elements $\langle ij| V | k l \rangle$ (refer to Appendix~\ref{sec:appendix_a}), which are defined as follows:
    \begin{equation}
        \langle ij|V|st\rangle = \delta _{l_{i} +l_{j} ,l_{s} +l_{t}}\int_{0}^{\infty } dq\ F_\alpha( q ) \vartheta _{i,s}( q) \vartheta _{j,t}(q),
        \label{eq:coulomb_integral_form}
    \end{equation}
Here, $\alpha$ denotes the dimensionless scale ratio $L/l_b$, where $l_b$ is the magnetic length, $\vartheta_{a,b}(q)$ are integral terms involving radial contributions, and $F_{\alpha}(q)$ is the form factor given by:
\begin{equation}
        F_\alpha(q) = \frac{\pi ^{2}\left[ 20\alpha ^{3} \pi ^{2} q^{3} +3\alpha ^{5} q^{5} +32\pi ^{4}\left( \alpha q+e^{-\alpha q} -1\right)\right]}{\alpha ^{2}\left( 4\pi ^{2} q+\alpha ^{2} q^{3}\right)^{2}}.
        \label{eq:formfactor}
\end{equation}
Finally, the last line of the Hamiltonian involves the light-matter interaction between the electron and hole pairs and the photonic mode confined inside the cavity with an amplitude of $g$. Using an exact diagonalization technique, we can investigate the structure of the excitons of the system and the negative trion resonances. Such properties can be achieved using a suitable bare-state basis to diagonalize the system Hamiltonian. Furthermore, owing to the conservation of total angular momentum, a block-diagonal approach can be employed to perform the diagonalization process in a fixed angular momentum subspace. For excitons, we use an electron-hole pair basis, represented by $| \Psi_X \rangle = |i \rangle_{e} \otimes | \overline{j} \rangle_{h}$. Conversely, negative trion states can be obtained by employing a set of states, $|\Psi_{T} \rangle = |i,j\rangle_{e} \otimes | \overline{k} \rangle_{h}$. This proposal accounts for the initial residual electron density in the semiconductor conduction band. For both cases, we constructed a set of states with a fixed total angular momentum. In addition, we established a cutoff criterion for single-particle states (see Appendix~\ref{sec:ordering}) based on the convergence of the first few excited states of the system.
\subsection{Trion-Exciton Polariton effective model}
Understanding the properties of a finite system is challenging because of the numerous energy levels involved in Hamiltonian formulation. However, under certain experimental conditions, it is possible to excite the system to produce exciton ground state (X) and negative trion (T) triplet bright states in the QW. To develop a simpler model that retains the Coulomb correlation information of matter quasiparticles, we can employ an effective Hamiltonian model to account for the interaction of X and T with the photons inside the cavity. Assuming a sufficiently large mean spatial separation between the X and T states, the Coulomb interaction between these states can be disregarded. In our theoretical approach, this can be achieved by computing the eigenstates corresponding to the exciton and trion ground states, using the Hamiltonian given by Eq.~(\ref{eq:sys_hamiltonian}). Consequently, we can express the matter-dressed states as
    \begin{eqnarray}
        |\mathrm{X}\rangle &=& \sum_{i,\overline{j}} \varphi_{\mathrm{X}}(i,\overline{j}) e^\dagger_{i} h^\dagger_{\overline{j}} |0 \rangle_{e} \otimes |0\rangle_{h}, \\
        |\mathrm{T}\rangle &=& \sum_{i,j, \overline{k}} \varphi_{\mathrm{T}}(i,j,\overline{k}) e^\dagger_{i} e^\dagger_{j} h^\dagger_{\overline{k}} |0\rangle_{e} \otimes| 0 \rangle_h. 
    \end{eqnarray}
Here, we denote $|0\rangle_{e,(h)}$ as the corresponding vacuum state for electrons (holes), and $\varphi_{\mathrm{X},(\mathrm{T})}$ as the linear combination coefficients of the Coulomb correlated states for the exciton and negative trion, respectively. Considering that the bright trion triplet has well-defined spin symmetry, this state decays to $|1\downarrow \rangle$ after the electron-hole combination through the recombination channel with left-polarized light. Therefore, we can describe the light-matter state basis spanned by $|\mathcal{\alpha} \rangle \otimes | \beta \rangle \otimes |n\rangle $ where $|n\rangle \in \{ |0\rangle, \cdots, |N\rangle\}$, $|\alpha \rangle \in \{ |\text{G} \rangle, |\text{X} \rangle \}$,  $| \beta \rangle \in \{ |1 \downarrow  \rangle, |\text{T} \rangle \}$, where $|1\downarrow \rangle$  is the electron state in the first orbital with a down $z$-spin component, and $n$ corresponds to the photon number. The effective Hamiltonian system can be represented as follows ($\hbar=1$).
\begin{eqnarray} \nonumber
        H_{\text{eff.}} &=& \omega a^\dagger a + \omega_{\mathrm{X}}
        \sigma^\dagger \sigma + (\omega_T - \omega_e) \tau^\dagger \tau  \\ \nonumber
        &+& g_{\mathrm{X}} \mathcal{P}_{\mathrm{X}}(B)( \sigma^\dagger a + a^\dagger \sigma ) + g_{\mathrm{T}} \mathcal{P}_{\mathrm{T}}(B)( \tau^\dagger a + a^\dagger \tau) \\
        &+& \Omega(t)( e^{-i\omega_L t }a^\dagger +  e^{i \omega_L t} a ),
        \label{eq:effective_hamiltonian}
\end{eqnarray} 
where $\omega$ is the photonic mode frequency. In addition, $\omega_{X}$, $\omega_{T}$, $\omega_{e}$ are the eigenfrequencies of the exciton, trion, and electron ground states, respectively. The raising and lowering operators for matter are defined as $\sigma = | \text{G} \rangle \langle \mathrm{X |} \otimes \mathbb{I}_{\mathrm{T}}\otimes \mathbb{I}_a$ and $ \tau = \mathbb{I}_{\mathrm{X}} \otimes |1 \downarrow \rangle \langle \mathrm{T} | \otimes \mathbb{I}_a $ which connects the excited state with the corresponding ground state of each quasiparticle. The effective Hamiltonian given by Eq. ~(\ref{eq:effective_hamiltonian}) includes terms such as the free energy of photons, exciton energy, and trion energy measured for the free electron state. Furthermore, the effects of the magnetic field on the light-matter interaction amplitudes were considered using dipole transition matrix elements~\cite{Kudlis2024}. This is because $\mathcal{P}_{\mathrm{X}}(B) = \langle 0|_{e}\otimes \langle 0 |_{h} h_{\overline{i} \downarrow} e_{i \uparrow } |\mathrm{X} \rangle $ and $\mathcal{P}_\mathrm{T}(B)= \langle 1 \downarrow| \otimes \langle 0 |_{h} h_{\overline{i} \downarrow} e_{i \uparrow } |\mathrm{T} \rangle$ together with the corresponding effective light-matter interaction constants $g_{\text{X}}$ and $g_{\text{T}}$. Note that, in our model, these effective interaction constants are considered different, as reported from experimental and theoretical perspectives ~\cite{Combescot2003, Rapaport2001}. Finally, the last term in the effective Hamiltonian refers to external coherent photonic pumping with frequency $\omega_L$ and time-dependent amplitude profile $\Omega(t)$ acting on the system.
\subsection{Optimization Algorithm for Quantum Gates}\label{sec:qg_synthesis}
\begin{figure}[!ht]
        \centering
        \includegraphics[width=0.8\linewidth]{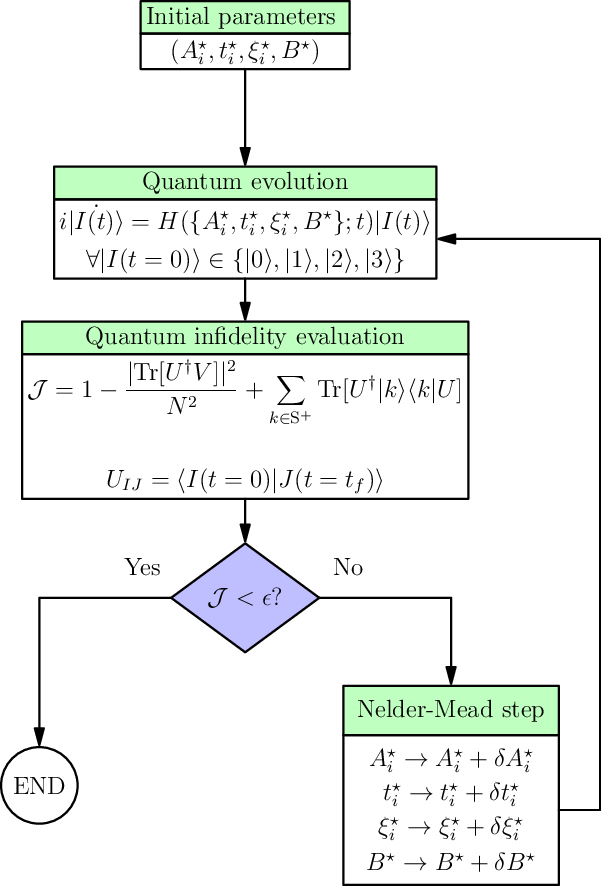}
        \caption{The flowchart illustrates the stages involved in the optimization algorithm for a quantum gate. For the interpretation of the colors in the figure(s), the reader is referred to the web version of this article.}\label{fig:figure-2}
\end{figure}

In this subsection, we consider the effective Hamiltonian to explore how the magnetic field and external coherent photonic pumping can be used to perform controlled quantum operations on the system. To achieve this, we restrict the Hilbert space of our effective model to a four-qudit-state basis using the following state mapping: $|\text{G},1\downarrow, 0 \rangle \rightarrow |0\rangle$, $|\text{G}, \text{T},0 \rangle \rightarrow |1\rangle$, $ |\text{X}, 1\downarrow, 0 \rangle \rightarrow |2\rangle $, $|\text{X}, \text{T}, 0 \rangle \rightarrow |3\rangle$. Additionally, a quantum control method was applied to determine the best set of parameters for external pulses. More precisely, for external pulses given by a train of Gaussian pulses, is defined as
\begin{equation}
    \Omega(t) = \sum_{i=1}^M A_i \exp\left( -\frac{(t-t_i)^2}{2\xi_i^2} \right),
\end{equation}
where parameters such as $A_i$, $t_i$ and $\xi_i$ denote the amplitude, center, and width of the $i-$th Gaussian pulse, respectively. The Fig.~\ref{fig:figure-2} shows a flowchart of the optimization algorithm used to obtain the quantum gate. Basically, each qudit bare basis state $|I\rangle \in \{|0\rangle, |1\rangle, |2\rangle, |3\rangle \}$ is numerically evolved using the time-dependent Schr\"odinger's equation, with an initial arbitrary set of parameters for the external pulse and magnetic field in the effective Hamiltonian model. The time-evolution operator is then obtained by projecting the final states onto their corresponding initial states, expressed as $U_{IJ} = \langle I(t=0) | J(t = t_f) \rangle$, where $|I(t)\rangle$ represents the state of a two-qubit basis at time $t$. We evaluated the quantum infidelity function $\mathcal{J} = 1 - |\text{Tr}[U^\dagger V ]|^2/N^2$, where $V$ is the target quantum gate and $N$ is the number of quantum states involved in the operation (two qubits and four qudits). 
It is important to note that we incorporate a penalty term to reduce the presence of states $k \in \mathrm{S}^+$, where $\mathrm{S}^{+}$ represents states that are not part of the qudit. The \textit{Nelder-Mead}~\cite{Nelder1965} multidimensional optimization algorithm is employed to iteratively update the pumping parameters and magnetic field to minimize the quantum infidelity $\mathcal{J}$.
\section{Results and discussion}\label{sec:results}
\subsection{Exciton and negative trion resonances}
\begin{figure}
    \centering
    \includegraphics[width=1.0\linewidth]{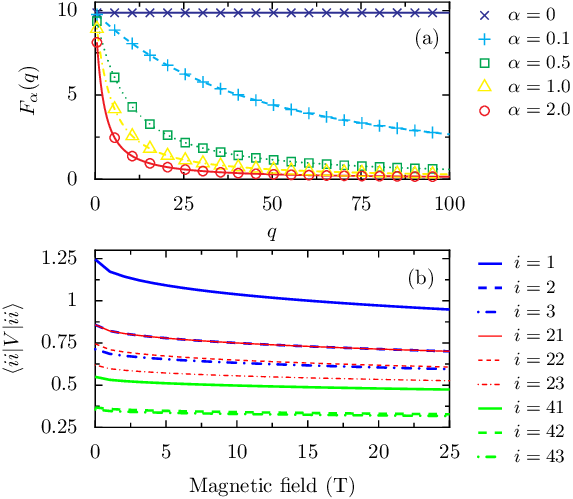}
    \caption{Panel (a) shows the form factor $F_\alpha(q)$ as a function of $q$ coordinates for different values of the $\alpha$ length ratio. The case of $\alpha = 0$ corresponds to the two-dimensional limit. Panel (b) shows the dependence of Coulomb matrix elements $\langle ii|V|ii\rangle$ on the external magnetic field. We considered the first three single-particle states at each level. Label $i$ refers to the corresponding Landau-level basis ordering. For the interpretation of the colors in the figure(s), the reader is referred to the web version of this article.
    }\label{fig:figure-3}
\end{figure}

\begin{figure}[!ht]
\centering
\includegraphics[width = 0.9\linewidth]{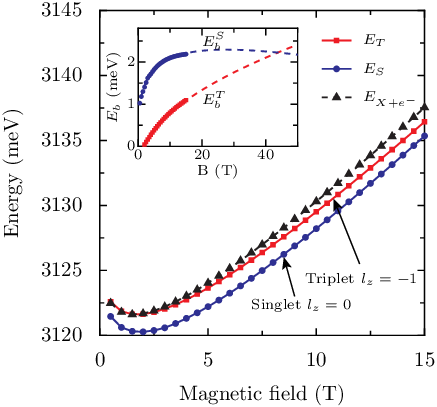}
\caption{Lowest-energy states of the negative trion in a GaAs QW as a function of the magnetic field. The main panel shows the triplet $E_T$, singlet $E_S$, and exciton plus free electron energies $E_{X+e^{-}}$. The inset shows the numerical calculations for the binding energies $E_b^{S}$ (blue circle line) and $E_b^{T}$ (red circle line) of the negative trion states relative to the exciton ground state and the free electrons as a function of the magnetic field. Extrapolation of the numerical results for the binding energies $E_b^{S}$ and $E_b^{T}$ are shown as blue and red dashed lines, respectively. A crossing between these two energies is predicted at $B\approx42T$. For the interpretation of the colors in the figure(s), the reader is referred to the web version of this article.}\label{fig:figure-4}
\end{figure}

In the following, we consider the typical parameter values reported for GaAs QW systems in our numerical simulations~\cite{PhysRevB.63.205324}. More precisely, $E_g = 1500~\text{meV}$, $\beta = 2.89 \sqrt{B}$  meV, $L = 10$ nm, $m_e = 0.0665m_0$, $m_h = 0.235 m_0$,  $g_e(B) = -0.01667 + 0.0052B$,  $g_h(B) = -0.05B$, $g_{\mathrm{X}} = 3$ meV, $g_{\mathrm{T}} = 0.5$ meV and we begin evaluating the impact on the Coulomb interaction when the magnetic field is changed through the form factor given by Eq.~(\ref{eq:formfactor}). Fig.~\ref{fig:figure-3}(a) shows the form factor distribution as a function of $q$ coordinates with different values of the characteristic length scale ratio $\alpha$. In general, it was found that the form factor decreased significantly, whereas $\alpha$ increased because of the confinement effect exerted by the magnetic field on the particles. Consequently,  Coulomb matrix elements are attenuated. Therefore, the form factor behaves as an amplitude that modulates the contributions of the radial function integrals. It is worth noting that $\alpha \to 0$ recovers the two-dimensional limit, in agreement with other theoretical results~\cite{HAWRYLAK1993475,PIETILAINEN1993809, Chakraborty1994}. 
Further analysis of the Coulomb matrix elements is presented in Fig.\ref{fig:figure-3}(b). In particular, we show Coulomb matrix elements of the form $\langle ii|V |ii\rangle$ as a function of the magnetic field for the first three single-particle states at each Landau level. Interestingly, we observed that the matrix elements decreased, in agreement with the behavior of the form factor, as shown in Fig.\ref{fig:figure-3}(a). However, it is worth noting that the potential energy of the system increased as the magnetic field increased because the Hamiltonian (see Eq. ~(\ref{eq:sys_hamiltonian})) incorporates Coulomb strength through $\beta$ term. The Fig.~\ref{fig:figure-4} shows the ground-state energy for the negative trion in subspaces with angular momentum $l_z=0$  (singlet state with energy $E_{S}$) and $l_z= -1$ (triplet state with energy $E_{T}$), as well as the ground-state energy for the exciton plus a free electron ($E_{X+e^{-}}$) in the QW system. We observed that at low magnetic fields, the Coulomb interaction energy is comparable to the cyclotron energy, as is well known in semiconductor physics~\cite{MacDonald1986}. In contrast, at high values of the magnetic field, it was found that the cyclotron energy dominates the Coulomb interaction energy; therefore, the system becomes uncoupled with an energy spectrum according to the Landau energy levels. The inset shows the trion binding energies $E_b^{S(T)} = E_{S(T)} - E_{X+e^{-}}$ for a single or triplet state as a function of the applied magnetic field. We observe that both energies increase monotonically, in complete agreement with experimental observations~\cite{BarJoseph2005,PhysRevB.56.15185,Riva2001}. Additionally, we extrapolated the numerical results for the binding energies $E_b^{S}$ (blue dashed line) and $E_b^{T}$ (red dashed line) in the high magnetic field regime. Interestingly, we found that our proposed model captures important features, such as the crossing between the binding energies at $B\approx42T$, which predicts a change in the negative trion ground state from the singlet to triplet state. This phenomenon has been reported both experimentally ~\cite{BarJoseph2005} and theoretically ~\cite{Redliński2002}.
\subsection{Single-qubit quantum gates}
\begin{figure}[!bht]
    \centering
    \includegraphics[width=0.9\linewidth]{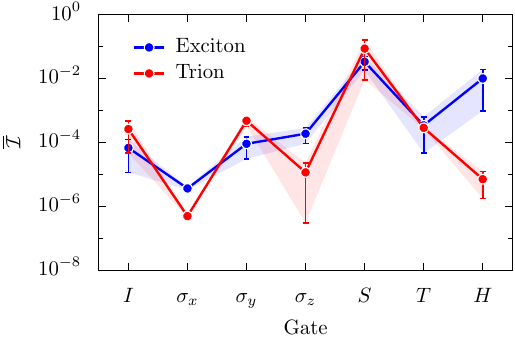}
    \caption{Numerical results for the quantum average infidelity using the optimization technique for creating single-qubit quantum gates for excitons (depicted by the blue circle line) and trions (represented by the red circle line). For the interpretation of the colors in the figure(s), the reader is referred to the web version of this article.
    }\label{fig:figure-5}
\end{figure}

\begin{figure}[!ht]
    \centering
    \includegraphics[width=0.9\linewidth]{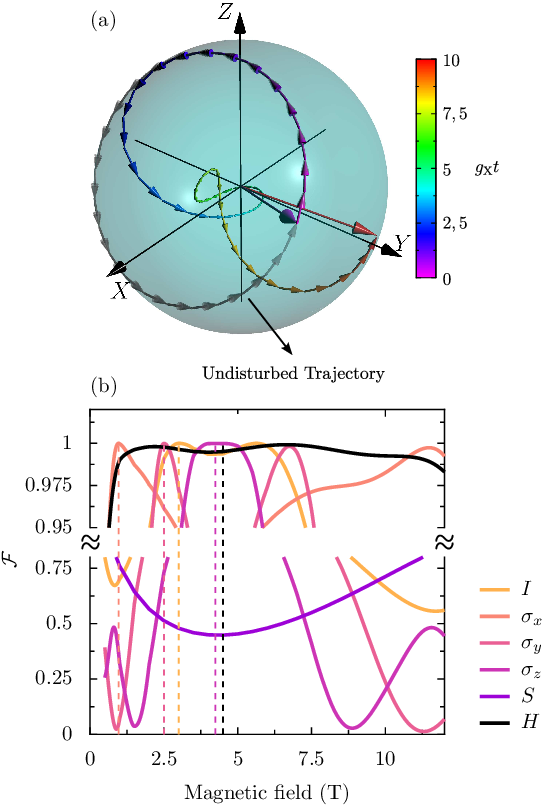}
    \caption{Panel (a) shows the time evolution of the initial state $|\psi_0 \rangle = \cos(\pi/5)|\mathrm{G},1\rangle + \sin(\pi/5)e^{i \pi/3} | \mathrm{X}, 0 \rangle $ on the modified Bloch sphere for obtaining the optimal Pauli-$Y$ gate. The undisturbed trajectory corresponds to the time evolution of the initial condition with an optimal magnetic field $B=2.51$ without photonic pumping. Panel (b) shows fidelity as a function of the external magnetic field. The vertical dashed lines indicate the optimal magnetic field for each gate. For the interpretation of the colors in the figure(s), the reader is referred to the web version of this article.}
    \label{fig:figure-6}
\end{figure}

The following demonstrates how photonic pulses and an external magnetic field can be used for the gate synthesis. More precisely, we applied the optimization method described in Section~\ref{sec:qg_synthesis} to determine a set of optimal parameters $A_i$, $t_i$ and $\xi_i$, as well as the external magnetic field $B$ that tunes the system evolution in a specific quantum gate. To achieve this, we label the qubit states as $|\text{G},1\rangle \rightarrow |0\rangle, |\text{X},0 \rangle \rightarrow |1\rangle$ for the exciton, and $|1\downarrow,1\rangle \rightarrow |0\rangle, |\text{T},0 \rangle \rightarrow |1\rangle$ for the trion. The quantum average infidelity was also considered using the following expression:
\begin{equation}
        \overline{\mathcal{I}} = 1 - \frac{1}{N} \sum_{n=1}^{N} |\langle \psi_n| U^\dagger V |\psi_n\rangle|^2.
    \end{equation}
The Fig.~\ref{fig:figure-5} shows the quantum average infidelity of the single-qubit gate synthesis process. Here, we computed the quantum average infidelity over $N=1000$ random initial states $|\psi_n \rangle = \cos( \theta_n/2 ) |0\rangle + \exp(i\phi_n)\sin(\theta_n/2)|1\rangle$ where $\theta_n \in [0,\pi]$ and $\phi_n \in [0,2\pi]$. It is noteworthy that $\langle \psi_n |U^\dagger$ corresponds to the quantum evolution of the initial state under the Hamiltonian given by Eq.~(\ref{eq:effective_hamiltonian}) and the optimal parameters obtained using the optimization algorithm. The term $V|\psi_n \rangle$ corresponds to the evolution of the initial state through the target quantum gate. In particular, we consider the maximum number of pumping pulses $M=5$ to study the following gates: identity ($I$), Pauli-$X$ ($\sigma_x$), Pauli-$Y$ ($\sigma_y$), Pauli-$Z$ ($\sigma_{z}$), square root of $Z$ ($S$), $\pi/8$ gate ($T$), and Hadamard gate ($H$). 
It should be noted that in both cases, quantum average infidelity was close to zero. This implies that the action of the photonic pumping pulse and the external magnetic field is optimal for controlling the desired states. For example, an interesting situation occurs when a photonic pumping pulse is neglected. The Pauli-$X$ gate can be obtained by transforming the effective Hamiltonian (Eq.~\ref{eq:effective_hamiltonian})) to Pauli matrix. In this representation, the temporal evolution of the state is determined by rotation around a unitary vector on a modified Bloch sphere. This vector can be written as:
    \begin{equation}
        \mathbf{n} = \frac{1}{ \sqrt{ G^2_{\mathrm{X},(\mathrm{T})} + (\delta_{\mathrm{X},(\mathrm{T})}/ 2)^2 }} \left( G_{\mathrm{X},(\mathrm{T})}, 0, \frac{\delta_{\mathrm{X},(\mathrm{T})}}{2} \right),
    \end{equation}
where $G_{\mathrm{X},(\mathrm{T})} = g_{\mathrm{X},(\mathrm{T})} \mathcal{P}_{\mathrm{X},(\mathrm{T})}(B)$ is the light matter amplitude and $\delta_{\mathrm{X},(\mathrm{T})} = \omega_{\mathrm{X(T)}}- \omega_{C}$ is the energy detuning between the light and matter. The unitary vector $\mathbf{n}$ living in the $XZ$plane can be reoriented by applying an optimized magnetic field. Thus, the system reaches the resonance condition, inducing a dynamic rotation around the $X$axis. However, in order to 
produce rotations around an arbitrary orientation on the modified Bloch sphere, the system must be perturbed by an external photonic pumping pulse. The general situation is illustrated in Fig.\ref{fig:figure-6}(a), where we consider as the arbitrary initial state $|\psi_0\rangle = \cos(\pi/5)|0\rangle + e^{i\pi/3} \sin(\pi/5)|1\rangle$ and the time evolution on the modified Bloch sphere is tracked~\cite{comment1}. When the photonic pumping pulse is neglected and the optimized magnetic field takes the value  $B = 2.51$T, a circular trajectory (undisturbed trajectory) on the modified Bloch sphere is registered around a unitary vector on the $XZ$ plane. Conversely, when the optimal photonic pumping pulse was considered, the registered trajectory on the modified Bloch sphere became significantly more complex. Initially, the trajectory followed the corresponding rotation as if it had no photonic pumping pulse; then, the trajectory was obtained in the modified Bloch sphere. This behavior is a consequence of population transfer to other states on a computational basis. Finally, the trajectory reached the surface of the modified Bloch sphere, successfully producing a rotation on $Y$-axis. Fig.~\ref{fig:figure-6}(b) shows the quantum fidelity as a function of the external magnetic field $\mathcal{F} = |\langle \psi_0| U^\dagger(B) V | \psi_0 \rangle |^2$. It was found that, in general, the fidelity approaches one for a limited region of the applied magnetic field, implying that is very sensitive to changes in the magnetic field. Except for Hadamard or Pauli-$Z$ gates, which are more robust to changes in the magnetic field because the quantum fidelity is approximately one for a wide range of magnetic field values. Similar results, not shown here, were obtained in gate synthesis when the trion-photon contributions were considered~\cite{comment2}. 
\subsection{Qudit quantum gate}
\begin{figure}[!ht]
    \centering
    \includegraphics[width=1.0\linewidth]{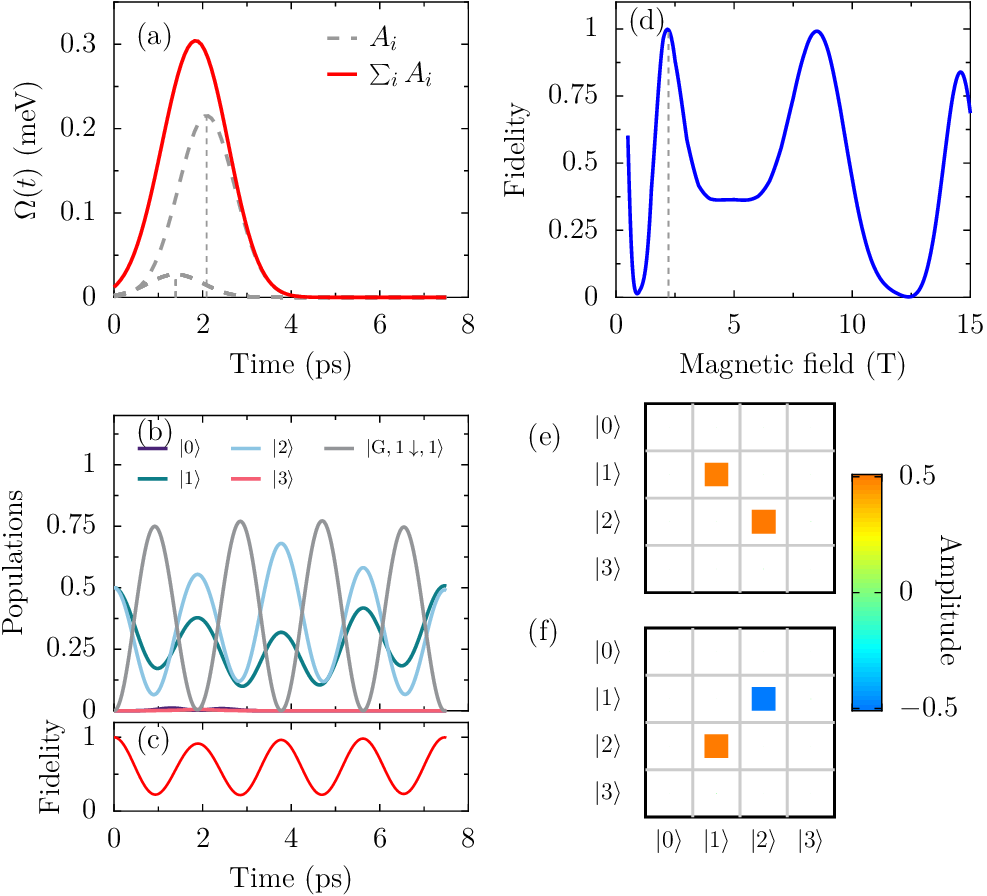}
    \caption{Panel (a) shows the optimal amplitude profile of a photonic pumping pulse. Panel (b) shows the time evolution of the qudit populations with the initial condition $|\phi_0\rangle = \frac{1}{\sqrt{2}}(|1\rangle + i|2\rangle) $ together with all the optimal parameters $(A_i,t_i,\xi_i, B)$ for the corresponding \textit{iSWAP} gate. Panel (c) shows the fidelity as a function of time. Panel (d) shows the fidelity as a function of the external magnetic field together with the optimal photonic pumping pulse. The vertical dashed line indicates the optimal magnetic field for which quantum fidelity is maximized. Panels (e)–(f) show the real and imaginary parts of quantum state tomography for the time-evolved state $U|\phi_0\rangle$. For the interpretation of the colors in the figure(s), the reader is referred to the web version of this article.}
    \label{fig:figure-7}
\end{figure}

\begin{figure}[!htb]
        \centering 
    \includegraphics[width=0.9\linewidth]{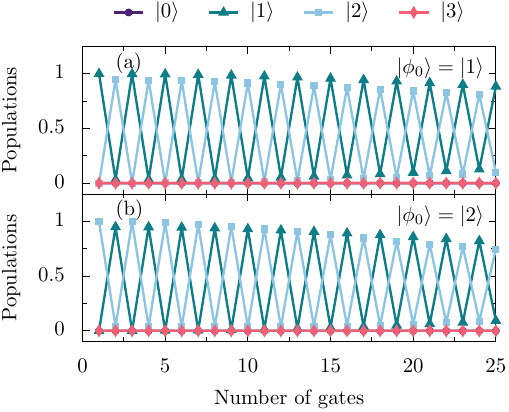}
        \caption{Panels (a)–(b) show the performance of the \textit{iSWAP} synthesized gate through numerical calculations of the populations for the quantum states $|1\rangle$ and $|2\rangle$ along each application of the $U$ operator. For the interpretation of the colors in the figure(s), the reader is referred to the web version of this article.
        }
        \label{fig:figure-8}
    \end{figure}

We numerically investigated the production of the \textit{iSWAP} four-level-qudit gate by controlling the detunings $\delta_{\text{X}}$ and $\delta_{\text{T}}$ and the modulation terms of the light-matter amplitudes $g_{\text{X}}\mathcal{P}_{\mathrm{X}}(B), g_{\text{T}}\mathcal{P}_{\mathrm{T}}(B)$ through the magnetic field and photonic pumping pulse. Fig.~\ref{fig:figure-7}(a) shows the amplitude profile $\Omega(t)$ as a function of time. The amplitude profile shown by the solid red line corresponds to an asymmetric Gaussian shape acting on the system. This shape profile is the result of the optimization algorithm for each term in the train of Gaussian pulses, shown as gray dashed lines. Fig.~\ref{fig:figure-7}(b) shows the time evolution of the populations when the initial state is considered as $|\phi_0\rangle = \frac{1}{\sqrt{2}}( |1\rangle + i|2\rangle)$ together with the optimized amplitude profile mentioned above and the optimal magnetic field $B=2.22$T. We observe that the occupation probability of states $|1\rangle$ and $|2\rangle$ shows a coherent exchange behavior influenced by the occupation of the auxiliary state $|\text{G},1\downarrow, 1\rangle$, whereas the states $|0\rangle$ and $|3\rangle$ remain with negligible occupation probability. This fact evidences how control of qudit states can be achieved through the light-matter interaction with intermediary states. Furthermore, it was found that the time evolution of fidelity as a function of time between the target quantum gate and the dynamically synthesized gate exhibits quasi-periodic behavior, reaching a maximum of $0.996$ at $ t = 7.5$ ps, as shown in Fig.\ref{fig:figure-7}(c). However, as shown in Fig.\ref{fig:figure-7}(d) shows the fidelity as a function of the external magnetic field and it is found that it depends strongly on the applied magnetic field. There are values for which the fidelity takes values close to unity. For example,  $\mathcal{F}=0.996$ at $B=2.22$T and  $\mathcal{F}=0.991$ at $B=8.5$T, which confirms that the optimization procedure was successfully used to achieve the proposed quantum gate. Figs.~\ref{fig:figure-7}(e)–(f) show the real and imaginary parts of quantum state tomography for the time-evolved state $U|\phi_0 \rangle$ at $t=7.5$ps when the quantum gate is successfully achieved. At this time, the time-evolved state exhibits the expected quantum structure of the \textit{iSWAP} operation, because both the populations and phases of the state are recovered. It should be noted that the time evolution operator $U$ should be obtained through an optimization procedure. Additionally, we validated the performance of our method for the synthesized gate by successive applications of the time-evolution operator $U$ on the states $|1\rangle$ and $|2\rangle$~\cite{Xian_2020} as shown in Fig.\ref{fig:figure-8}(a)-(b). It is confirmed that the population interchanges between $|1\rangle$ and $|2\rangle$ states, as is well known in the \textit{iSWAP} gate. We observed a systematic reduction in the occupations during successive applications of the $U$ operator as a consequence of numerical errors in the optimization algorithm.
\section{ \label{sec:conclusion} Conclusions}
In this study, we introduced a theoretical control protocol to synthesize single-qubit and four-level quantum gates using the Hamiltonian dynamics a trion-exciton-polariton system in semiconductor QWs. Our approach leverages the effects of external electromagnetic fields on charge carriers within the active media of an optical cavity to manipulate the phase and amplitude of the encoded state populations. For qubits, we encoded the states into the lowest polariton, whereas for qudits, we selected the lowest matter states without photon contributions. We found that our methodology enables precise modeling of the Coulomb interactions between charge carriers and allows incorporation of the effects of the magnetic field on the system properties with high fidelity. We examined the behavior of the Coulomb matrix elements in the QW system as a function of the magnetic field, and found that the amplitude of each matrix element depends significantly on the behavior of the form factor $F_\alpha(q)$, resulting in blue shifts due to the applied magnetic field in the exciton and trion energies, as well as in their corresponding ground-state structures. It was also demonstrated that the fidelity is sensitive to magnetic field changes for each quantum-controlled gate, underscoring the importance of magnetic control in the synthesis process. In particular, we constructed an \textit{iSWAP} gate in a four-level qudit with $99.6\%$ fidelity, confirming that the light-matter interaction can be controlled by the magnetic field and allows population transfer between qudit states to achieve the desired interchange in gate operation.
\section*{Funding}
A.F.U.R., H.V.-P. and E.A.G. acknowledge funding from project "Ampliación del uso de la mecánica cuántica desde el punto de vista experimental y su relación con la teoría, generando desarrollos en tecnologías cuánticas útiles para metrología y computación cuántica a nivel nacional", BPIN 2022000100133, from SGR of MINCIENCIAS, Gobierno de Colombia.
\section*{Declaration of competing interest}
The authors declare that they have no known competing financial interests or personal relationships that could have influenced the work reported in this study.
\section*{Acknowledgements}
The authors acknowledge technical and computational support from Grupo de \'Optica e Informaci\'on Cu\'antica (GOIC-UNAL). A.F.U.R.,
H.V.-P and E.A.G. thanks financial support from project ‘Ampliación del uso de la mecánica cuántica desde el punto de vista experimental y su relación con la teoría, generando desarrollos en tecnologías cuánticas útiles para metrología y computación cuántica a nivel nacional’, BPIN 2022000100133 from SGR of MINCIENCIAS; Gobierno de Colombia. E.A.G. acknowledges the financial support from the Universidad del Quindío under Project No. 1141.
\section*{Author contributions}
\begin{itemize}
    \item A. F. Urquijo Rodríguez: Analyzed and interpreted the data; Contributed analysis tools and wrote the paper.
    \item E. A. Gómez González: Conceived and designed the analysis; Analyzed and interpreted the data; and wrote, reviewed and edited the manuscript.
    \item H. Vinck-Posada: Conceived and designed the analysis and analyzed and interpreted the data.
\end{itemize}
\appendix
\section{Computation of Coulomb matrix elements\label{sec:appendix_a}} 
\begin{figure}[ht!]
    \centering
    \includegraphics[width=0.9\linewidth]{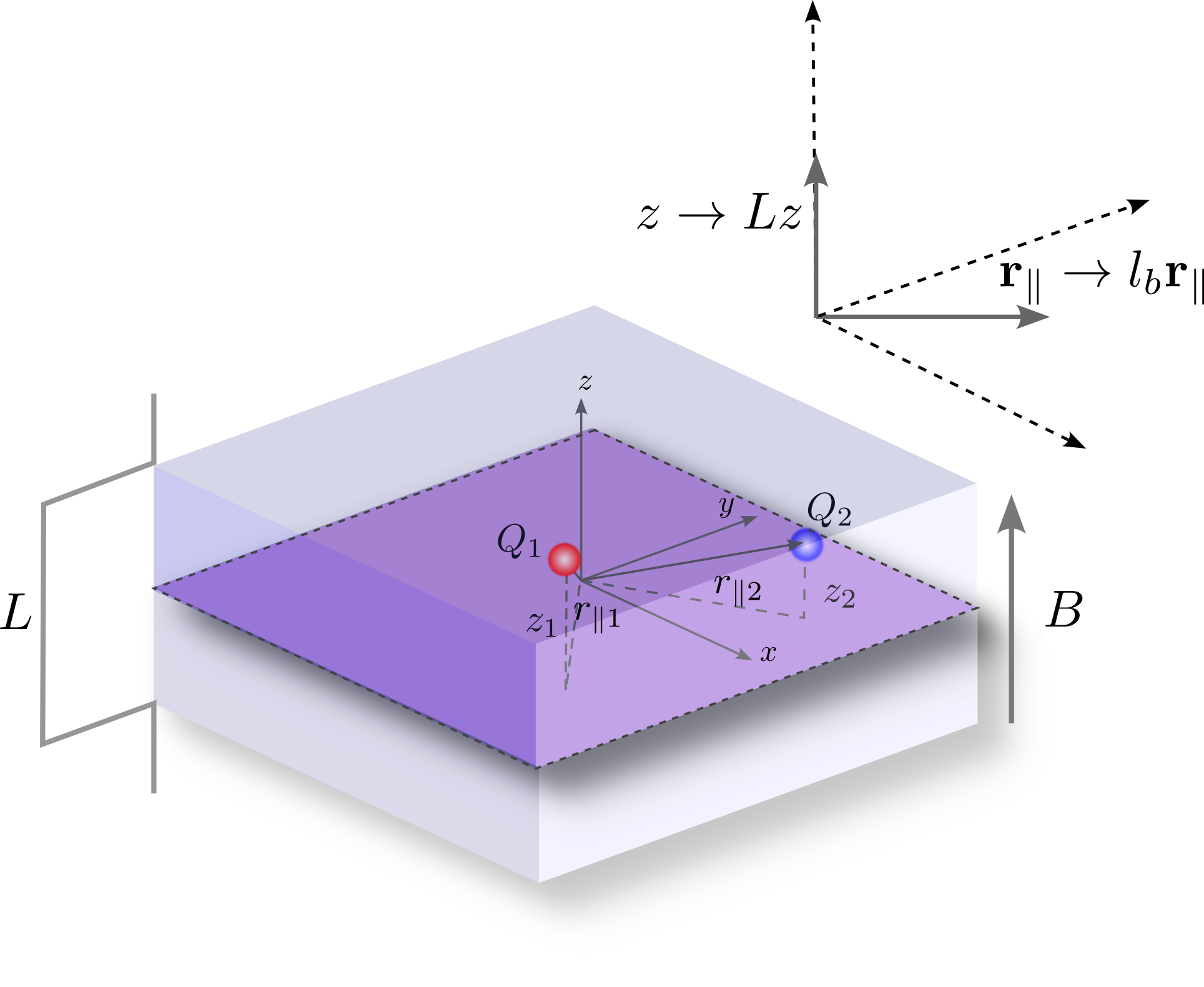}
    \caption{The schematic illustrates the geometry of the system, wherein the charged particles are confined within the QW under the influence of an external perpendicular magnetic field, denoted as $B$. The radial coordinate $r$ and transverse coordinate $z$ define the particle positions. The figure on the right shows the scaling of the dimensionless coordinates. For the interpretation of the colors in the figure(s), the reader is referred to the web version of this article.
    }
    \label{fig:figure-9}
\end{figure}

Let us consider the Hamiltonian that describes the electrostatic interaction of charge carriers inside a semiconductor QW, together with the presence of an external magnetic field. In particular, we introduce  dimensionless particle coordinates $(\mathbf{r}_{\parallel}, z)$, as shown in Fig.~\ref{fig:figure-9}. 
\begin{equation}
    V(\mathbf{r}_1-\mathbf{r}_2) = \frac{Q^2}{4\pi \epsilon \epsilon_0 l_b} \frac{1}{\sqrt{|\mathbf{r}_{\parallel1}-\mathbf{r}_{\parallel2}|^2+ \alpha^2(z_1-z_2)^2}},
    \label{eq:coulomb_potential}
\end{equation}
where $l_b=\sqrt{2\hbar/QB}$ denotes the in-plane magnetic length, $\alpha$ is defined as the ratio of the width of the QW to the magnetic length $L/l_b$, $Q$ is the fundamental electron charge, $B$ is the magnetic field intensity, $\epsilon$ is the relative permittivity of the material, and $\epsilon_0$ is the electrical permittivity of vacuum. Considering the second quantization framework, we choose as basis functions the product of the eigenfunctions of a single electron subject to an external perpendicular magnetic field to describe the in-plane problem and the ground state of an infinite symmetric QW in the transverse direction, as follows:
    \begin{eqnarray}\notag
        \phi_i(\mathbf{r}) &=& c_{n_i,l_i} r^{|l_i|} e^{-\frac{r^2}{2}} L_{n_i}^{|l_i|}(r^2) e^{i l_i \theta} \cos(\pi z) \\
        &=& c_{n_i,l_i} R_{n_i,l_i}(r) e^{i l_i \theta}\zeta(z).
    \end{eqnarray}
Here, $n_i$ and $l_i$ denote the radial and angular quantum numbers of the states $|i\rangle$ and $L_{n_i}^{|l_i|}(r)$ are the associated Laguerre polynomials. The normalization constant is given by $c_{n_i,l_i} = \sqrt{2n_i!/[\pi(n_i + |l_i|)!]}$. We assume the first QW sub-band approximation owing to the large energy gap between the first excited state and the QW ground state compared to the typical Coulomb strength in thin QWs. In addition, the spin degrees of freedom ($s_i$) were neglected because the influence of Coulomb interaction preserves their inherent properties. We define $\langle ij|V| st \rangle$ as the two-body Coulomb matrix element as
    \begin{equation}
        \langle ij|V|st\rangle =\int d^{3} r_1 d^{3} r_2 \phi _{i}^{*} (\mathbf{r}_1 )\phi_{j}^{*} (\mathbf{r}_2)V(\mathbf{r}_1 -\mathbf{r}_2 )\phi_{t} (\mathbf{r}_2 )\phi_{s} (\mathbf{r}_1).
    \end{equation}
Note that $\mathbf{r}$ is a 3D vector, and labels ($1,2$) denote the position vectors of each particle. By using the Coulomb potential, Eq.~(\ref{eq:coulomb_potential}) and single-particle wave functions using Fourier integrals~\cite{Chakraborty1994}:
    \begin{eqnarray}
        \phi_k(\mathbf{r}) &=& \frac{1}{(2 \pi)^3}\int d^3 q \psi_k(\mathbf{q}) e^{i \mathbf{q}\cdot \mathbf{r}}, \\
        V(\mathbf{r}_1-\mathbf{r}_2) &=&  \frac{1}{(2 \pi)^3} \int d^3 q \tilde{V}(\mathbf{q}) e^{i \mathbf{q}\cdot (\mathbf{r}_1 - \mathbf{r}_2 )},
    \end{eqnarray}
where $\psi_k(\mathbf{q}), \tilde{V}(\mathbf{q})$ correspond to the single-particle wave function and the Coulomb potential Fourier transform, respectively. The Coulomb matrix element takes the following form~\cite{Zaratiegui}
    \begin{eqnarray} \notag
        \langle ij|V|st\rangle &=& \frac{1}{( 2\pi )^{9}} \int d^{3} q \int d^{3} q_{1}\int d^{3} q_{2}\psi _{i}^{*}(\mathbf{q}_{1}) \psi_{j}^{*}(\mathbf{q}_{2})\\
        &\times& \psi_{t}(\mathbf{q}_{2} +\mathbf{q}) \psi _{s}(\mathbf{q}_{1} -\mathbf{q}) \tilde{V}(\mathbf{q}).
    \end{eqnarray}
It is straightforward to see that the last expression can be rewritten in a more compact form by grouping terms as follows:
    \begin{equation}
        \langle ij|V|st\rangle = \frac{1}{(2\pi)^3} \int d^3 q A_{i,s}(\mathbf{q}) B_{j,t}(\mathbf{q}) \tilde{V}(\mathbf{q}),
        \label{eq:qrepre}
    \end{equation}
where $A_{i,s}(\mathbf{q})$ and $B_{j,t}(\mathbf{q})$ are the convolution integrals given by
    \begin{eqnarray}
        A_{i,s}(\mathbf{q}) &=& \frac{1}{(2\pi)^3}\int d^3 q_1 \psi^*_i(\mathbf{q}_1)\psi_s(\mathbf{q}_1 - \mathbf{q}), \\
        B_{j,t}(\mathbf{q}) &=& \frac{1}{(2\pi)^3}\int d^3 q_2 \psi^*_j(\mathbf{q}_2)\psi_t(\mathbf{q}_2 + \mathbf{q}).
        \label{eq:convolution}
    \end{eqnarray}
Now, the previous integrals can be transformed into the space coordinate representation and together with the Fourier transform of the 3D dimensionless Coulomb potential:
    \begin{equation}
        \tilde{V}(\mathbf{q}) = \frac{4 \pi \alpha}{q_z^2 + \alpha^2 q^2}.
    \end{equation}
We obtain the final form of the Coulomb matrix element, which can be expressed as:
    \begin{equation}
        \langle ij|V|st\rangle = \delta _{l_{i} +l_{j} ,l_{s} +l_{t}}\int_{0}^{\infty } dq\ F_\alpha( q ) \vartheta _{i,s}( q) \vartheta _{j,t}( q),
        \label{eq:coulomb-integral}
    \end{equation}
with
    \begin{equation}
        F_\alpha(q) = \frac{\pi ^{2}\left[ 20\alpha ^{3} \pi ^{2} q^{3} +3\alpha ^{5} q^{5} +32\pi ^{4}\left( \alpha q+e^{-\alpha q} -1\right)\right]}{\alpha ^{2}\left( 4\pi ^{2} q+\alpha ^{2} q^{3}\right)^{2}},
        \label{eq:formfactor2}
    \end{equation}
and
    \begin{eqnarray}
        \notag
        \vartheta_{a,b}( q) &=& c_{n_a,l_a}c_{n_b,l_b} \int_{0}^{\infty } dr~r^{|l_{a} |+|l_{b} |+1} e^{-r^{2}} \\
        &\times& L_{n_{a}}^{|l_{a} |}\left( r^{2}\right) L_{n_{b}}^{|l_{b} |}\left( r^{2}\right) J_{|l_{a} -l_{b} |}( qr).
    \end{eqnarray}
Here, $J_{|l_a - l_b|}(qr)$ denotes the Bessel function of the first type. It is important to highlight the explicit conservation of angular momentum, as shown in Eq.~(\ref{eq:coulomb-integral}) because the scattering processes involved in the Coulomb interaction preserve the angular momentum of particles. Owing to the highly oscillatory nature of the functions $\eta_{a,b}(q)$ at elevated angular momentum values, Eq.~(\ref{eq:coulomb-integral}) was computed numerically using the quadrature method~\cite{Milovanovic2014} to ensure high precision. The finite structure of the Coulomb matrix element is significantly influenced by the form factor $F_\alpha(q)$, which is an envelope function that exhibits a magnetic-field-dependent decay rate. 
\section{Landau basis ordering~\label{sec:ordering}}
\begin{figure}[!bht]
        \centering
      \includegraphics[width=0.9\linewidth]{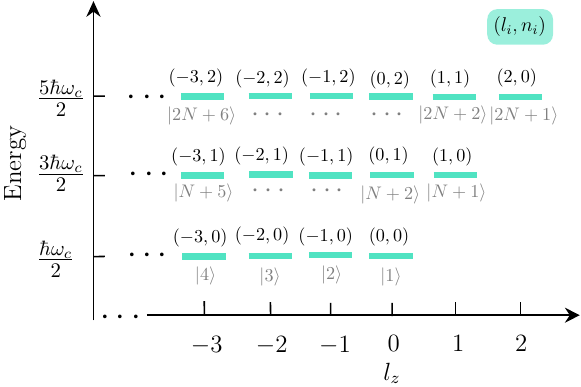}
        \caption{The diagram illustrates the ordering of the Landau basis, in which each layer comprises $N$ degenerate electronic states. The Dirac notation $|i\rangle$ denotes the ordering number, whereas the pair $(n_i, l_i)$ specifies the associated radial and angular quantum numbers. For the interpretation of the colors in the figure(s), the reader is referred to the web version of this article.}
        \label{fig:figure-10}
    \end{figure}

In this appendix, we elucidate the method by which the ordering of quantum states is utilized in all the calculations in this study. Specifically, Fig.~\ref{fig:figure-10} illustrates the basis for ordering Landau levels. We assume that the states are evenly distributed across the three levels, with $N$ states per level ($N = 20$ for the numerical simulations). Furthermore, each layer comprises degenerate electronic states, with the energy determined by (neglecting the spin) $\hbar \omega_c^{(e)}(B)/2\cdot( 2n_i + |l_i| - l_i + 1 )$, where $\omega_c(B)$ represents the electron cyclotron frequency. Each state is identified by the ordering number $|i\rangle$ and its corresponding quantum number $(n_i, l_i)$, which denote the radial and angular momentum numbers, respectively.
\bibliography{biblio}

\begin{thebibliography}{47}
\bibitem [{Annotation 1()}]{comment1}%
  \BibitemOpen
  Annotation 1,\ \href@noop {} {}\bibinfo {note} {The numerical results were
  obtained by considering the effective Hamiltonian as described in
  Eq.~(\ref{eq:effective_hamiltonian}) while disregarding the contributions
  from trions. Specifically, we omitted all terms that involve trion
  operators.}\BibitemShut {Stop}%
\bibitem [{Annotation 2()}]{comment2}%
  \BibitemOpen
  Annotation 2,\ \href@noop {} {}\bibinfo {note} {The numerical results for
  trion-photon contributions are obtained by considering the effective
  Hamiltonian as described in Eq.~(\ref{eq:effective_hamiltonian}) and neglect
  all terms that involve exciton operators.}\BibitemShut {Stop}%
\end{thebibliography}
\end{document}